# Vers l'auto-gestion d'un réseau de radio cognitive

Mohammed Zakarya Baba-Ahmed, Badr Benmammar
Dept. Électronique ET Électrotechnique
Laboratoire de Télécommunication de Tlemcen, LTT
Tlemcen, Algérie
{zaki.babaahmed, badr.benmammar}@gmail.com

Fethi Tarik Bendimerad
Dept. Électronique ET Électrotechnique
Laboratoire de Télécommunication de Tlemcen, LTT
Tlemcen, Algérie
ftbendimerad@gmail.com

*Résumé—* **La radio cognitive (RC) évolue dans des domaines aussi différents que variés, l'un de ces domaines applicatifs est les réseaux. Dans ce papier, nous proposons une nouvelle approche de la RC qui vise à gérer la transmission de la vidéo conférence d'un utilisateur secondaire (SU) pour différentes classes gérer par un utilisateur primaire (PU) en possession d'une licence sur le spectre et de traiter les éventuelles échecs du système et des applications informatiques par l'introduction d'un réseau autonome afin de rendre ces systèmes capables de s'autogérer avec un minimum d'intervention humaine.**

I. INTRODUCTION

L'idée de la radio cognitive a été présentée officiellement par Joseph Mitola III à un séminaire à KTH, l'institut royal de technologie, en 1998, publié plus tard dans un article de Mitola et Gerald Q. Maguire en 1999[1].

Le terme de radio cognitive a été fréquemment utilisé pour parler d'un système capable de prendre conscience de son environnement et de tirer profit de cette information. Parfois, il est considéré de façon plus restrictive comme un système disposant d'une grande agilité en fréquence pour explorer les opportunités qui peuvent exister dans le spectre fréquentiel [2].

La radio cognitive permet une bonne gestion du spectre en occupant ou bien en exploitant les bandes inoccupées des spectres radio, et bien sur permettant ainsi d'améliorer la gestion du spectre. Cela ce fait grâce à la radio logiciel restreinte SDR (Software Defined Radio).

La SDR est un système de communication radio qui peut s'adapter à n'importe quelle bande de fréquence et recevoir n'importe quelle modulation en utilisant le même matériel [1].

Les réseaux autonomes sont principalement représentés par l'auto-gestion qui vise à rendre les systèmes informatiques moins dépendants des utilisateurs. Cette dernière est caractérisée par quatre points bien distincts: L'auto-optimisation, l'auto-configuration, l'auto-protection ainsi que l'auto-guérison, comme indiqué dans la figure 1.

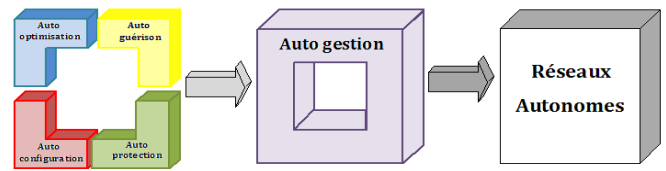

Figure 1. Les quatre éléments de base

Un système autonome ne sera jamais régler pour la situation présente. Il sera constamment suivi des objectifs du système prédéfinis ou des niveaux de performances pour s'assurer que tous les systèmes sont en cours d'exécution à un niveau optimal, il doit être en mesure d'installer et de configurer le logiciel automatiquement. Il doit aussi identifier, détecter, et protéger les précieuses ressources d'entreprise à partir de nombreuses menaces, comme il aura la possibilité de découvrir et de réparer les éventuels problèmes afin de s'assurer que les systèmes fonctionnent correctement [3].

Dans ce papier nous commenceront par établir le lien entre la radio cognitive et les réseaux autonomes par l'intégration des caractéristiques de l'auto-gestion dans les systèmes cognitifs, ensuite nous allons présenter une approche qui permet de gérer les éventuels problèmes d'interférences susceptible d'apparaître entre un utilisateur primaire possédant une licence sur le spectre et un utilisateur secondaire de la RC qui va allouer des canaux sur ce spectre.

II. L'AUTONOMIE DANS LE CONTEXTE DE LA RADIO COGNITIVE

L'autonomie dans les réseaux de radio cognitive est principalement focaliser sur la gestion du spectre qui permet l'amélioration du débit sans pour autant à dégrader les communications des autres. Plusieurs travaux en étaient présentés sur les différentes caractéristiques de l'auto-gestion:

A. *Auto-optimisation du moteur cognitif*

Ce travaille a été déjà réalisé précédemment par [4] qui permet l'adaptation totale et autonome du moteur cognitif en :

- Respectons le cadre de régulation qui contrôle l'accès au spectre.

- Satisfaire les besoins de l'utilisateur en termes de qualité de service.

- Assurons une gestion optimisée des ressources disponibles [4].

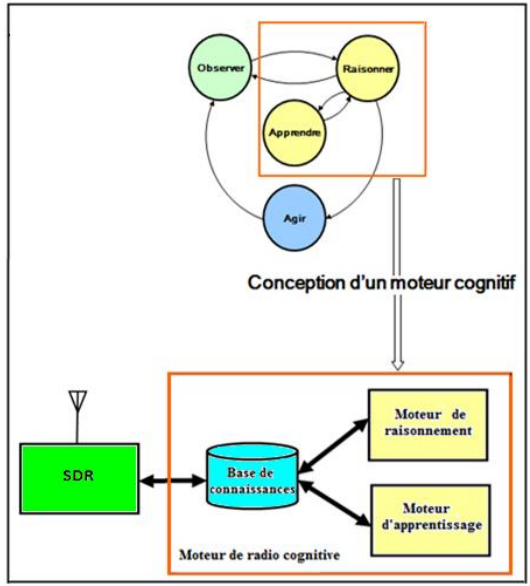

Figure 2. Auto-optimisation du moteur cognitif

Dans la figure 2, les éléments de la RC communiquent avec le support SDR à travers le moteur cognitif. Ce dernier représente la partie chargée de l'optimisation ou du contrôle du module SDR en se basant sur quelques paramètres d'entrée tels que les informations issues de la perception sensorielle pour le raisonnement du système ou de l'apprentissage de l'environnement radio, du contexte utilisateur, et de l'état du réseau.

Ici, la base de connaissances maintient les états du système et les actions disponibles. Le moteur de raisonnement utilise la base de connaissances pour choisir la meilleure action. Le moteur d'apprentissage effectue la manipulation des connaissances basées sur l'information observée (des informations sur la disponibilité des canaux, le taux d'erreurs dans le canal, etc.) [1].

B. *Algorithme d'auto-configuration de la couche 2*

Dans un réseau comprenant des dispositifs compatibles, l'auto-configuration de la couche 2 dans la RC implique la détermination d'un ensemble commun de canaux afin de faciliter la communication entre les nœuds participants. C'est un défi unique, car les nœuds du réseau RC ignorent peut-être :

- Leurs voisins ;

- Les canaux sur lesquels ils peuvent communiquer avec un voisin.

Les auteurs de [5] ont proposé un algorithme distribué de temps efficace pour l'auto-configuration de la couche 2 d'un réseau RC.

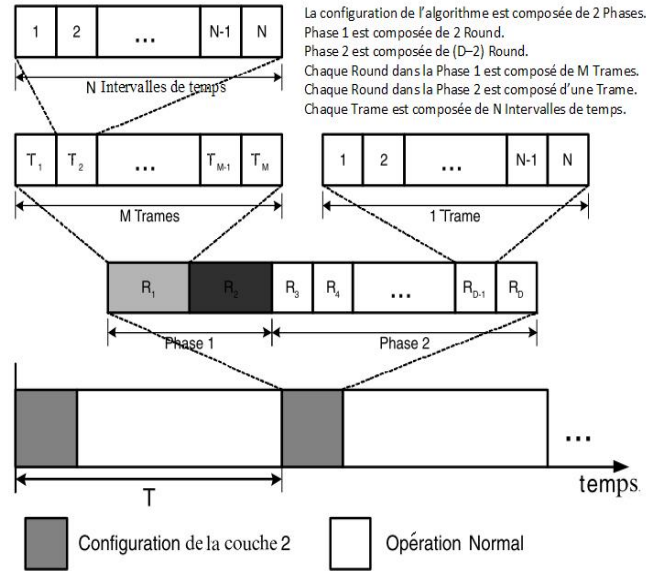

Figure 3. Cycle de fonctionnement d'un nœud radio cognitif [5]

Au cours du processus d'auto-configuration de la couche 2, le système à accès multiple par répartition dans le temps (TDMA) est utilisé pour la communication entre les nœuds. Le temps est divisé en O(D) Rounds. Un round est défini comme le temps nécessaire pour que chaque nœud communique avec ses voisins en utilisant un mécanisme de diffusion (locale). Chaque round se compose d'intervalles de tailles égales appelées Trames. Le nombre de trames dans un round peut varier, comme indiqué dans la Figure 3. Un round dans la phase 1 se compose de m trames, un round dans la phase 2 consiste en une seule trame. Chaque trame est divisée en N intervalles de temps, chacun de longueur égale. Nœud i transmet pendant l'iéme intervalle de temps dans chaque trame (voir Figure. 3) et tous les autres nœuds sont en mode de réception pendant l'iéme intervalle de temps [5].

C. *Auto-conscience dans le cycle de cognition*

L'auto-conscience démontre le développement de concepts qui constituent l'un des principaux problèmes dans la conception des réseaux cognitifs en auto-gestion pour la vision d'Internet du futur tel qu'il est actuellement développé dans le projet Self-NET [6].

La figure 4 unifie et formule quelques-unes des principales questions nécessaires au développement des systèmes cognitifs en auto-conscience et effectue actuellement la mise en œuvre de ces derniers avec des bancs d'essai réels qui révéleront la réalisation pratique et des problèmes de coordination dans le déploiement des cycles cognitifs pour les cas d'utilisation et l'application du cadre théorique présenté dans la figure 4.

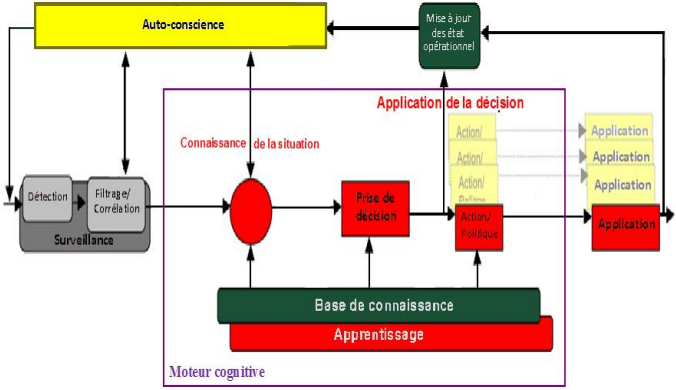

Figure 4.  Architecture logique Self-NET d'un cycle cognitif pour l'auto-conscience [6]

### III. SOLUTION PROPOSÉE

Nous nous intéressons dans ce papier à deux caractéristiques des réseaux autonomes, à savoir l'auto-protection et l'auto-guérison. On s'intéresse également à une application critique en termes de Qualité de Service (QoS) qui est la vidéo conférence. Donc notre objectif est de satisfaire ce type d'application étant donné un réseau autonome basé sur la RC.

Pour avoir une bonne QoS de la vidéo conférence, il est nécessaire que [7] :

- Le débit (bande passante) doit être > 384 Kb / s.
- Le délai doit être < 200 ms.
- La gigue doit être < 30 ms.
- Le taux d'erreur doit être < 1%.

TABLE I.  TROIS CLASSES POUR LA QOS

| Caractéristiques | Classes | | |
|---|---|---|---|
| | *C1* | *C2* | *C3* |
| Bande passante | > 384 Kb/s | Entre 162 et 384 Kb/s | < 162 Kb/s |
| Délai | < 200 ms | Entre 200 et 400 ms | > 400 ms |
| Gigue | < 30 ms | Entre 30 et 60 ms | > 60 ms |
| Taux d'erreur | < 1 % | 1 % | > 1 % |

Nous définissons dans le tableau 1, trois classes de QoS, la classe C1 est l'idéal pour un utilisateur RC car elle offre les meilleurs caractéristiques en termes des 4 paramètres. La classe C2 est moyenne en termes de QoS par contre la classe C3 est la plus mauvaise en termes de QoS.

#### A. L'architecture proposée

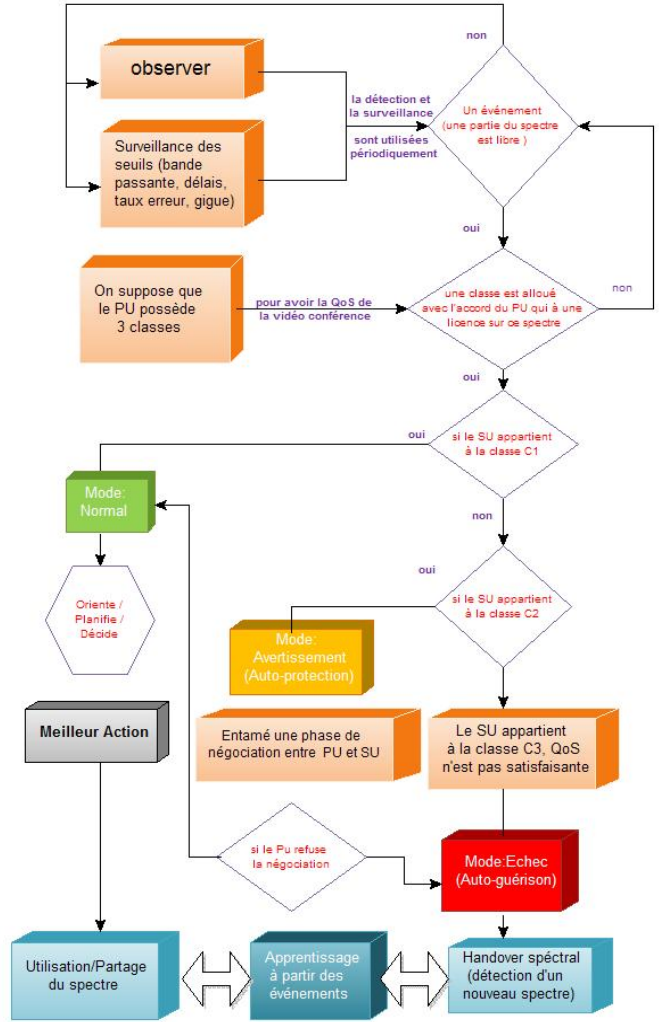

Figure 5.  Auto-protection et auto-guérison d'un nœud radio cognitif pour la gestion des échecs

#### B. Principe

Le principe de cette approche consiste qu'un utilisateur secondaire (SU) va détecter et surveiller les seuils de la QoS périodiquement jusqu'à qu'il trouve une partie libre d'un spectre sous licence gérée par un utilisateur primaire (PU).

Trois scénarios sont possibles dans ce cas :

1/ Si le PU décide d'allouer une partie du spectre assurant au SU une classe C1, donc le SU est en mode *Normal*. Il oriente, planifie et décide de la meilleur action à exécuter puis il utilise/partage le spectre avec le PU.

2/ Si le PU décide d'allouer une partie du spectre assurant au SU une classe C2, dans ce cas, on est en mode *Avertissement*, l'utilisation de cette classe (caractéristiques à condition moyenne) peut provoquer des dégradations du signal lors de la transmission, ce qui va sérieusement déranger l'utilisateur secondaire puisqu'il est sensé assurer la QoS de la vidéo conférence, alors une phase de négociation sera entamé entre le PU et le SU, c'est ce que nous appelons *l'Auto-protection*. Après négociation nous aurions deux cas possibles :

- Le premier cas sera produit si le PU coopère avec le SU et accepte d'assurer la classe C1 pour la transmission, donc le SU va revenir au premier scénario (c.à.d. en mode *Normal*).
- Le deuxième cas se produira si le PU refuse tout type de négociation, le SU va donc entamer une nouvelle phase qui est mentionné dans le troisième scénario.

3/ Si le PU décide d'allouer une partie du spectre assurant au SU une classe C3, dans ce cas la, on est dans le mode *Échec*, le signal sera très faible et l'échec de connexion sera garantie et aucune phase de négociation ne sera accordé, donc une nouvelle phase est exigée, on parle alors de *l'Auto-guérison*. Dans cette phase nous allons effectués un changement du spectre ou autrement dit un *Handover Spectrale*.

L'Auto-protection est engendrée par la phase de négociation qui sollicitera un accès dynamique au spectre, les différentes approches existantes pour l'allocation dynamique du spectre sont :

- MAC (Medium Access Control) ;
- Négociation locale ;
- Théorie des jeux ;
- SMA (Systèmes Multi Agents) ;
- Chaînes de MARKOV.

Cette dernière approche sera la plus favorable à la négociation de notre travail car normalement dans les approches les plus connues basées sur les chaînes de MARKOV, les auteurs modélisent les interactions entre les utilisateurs et calculent les probabilités de blocage et non-achèvement comme principaux paramètres d'évaluation [8].

L'Auto-guérison est engendrée quand à elle par la phase de changement du spectre (*Handover spectrale*) qui effectuera une nouvelle détection d'un autre spectre avec les mêmes exigences en terme de QoS pour la vidéo conférence.

L'apprentissage se fait par les événements qui se sont déroulés au cours des trois modes et après les résultats obtenus par la gestion des deux phases, ce qui amène à enrichir la base de connaissance par un *Apprentissage automatique*.

IV. CONCLUSION

Dans ce papier nous avons présenté plusieurs parties de l'auto-gestion reliant la radio cognitive avec les réseaux autonomes mais notre travail a été principalement focalisé sur la transmission de la vidéo conférence par l'intégration de deux parties de l'autonomie à savoir l'auto-protection et l'auto-guérison, avec le recours à la négociation en cas de menace d'interférence entre le PU et le SU ou le cas de changement du spectre causé par une rupture de coopération.

Dans nos futurs travaux, nous allons chercher à minimiser le taux d'échec de négociation entre les PUs et le SUs coexistant dans le même spectre et mesurer l'importance de l'auto-gestion dans les réseaux de radio cognitive assurant des contraintes applicatifs pour des applications critiques comme la vidéo conférence.